# E-learning Information Technology Based on an Ontology Driven Learning Engine

Liskin Viacheslav
Research student, Department of Applied Mathematics
National Technical University of Ukraine "Igor Sikorsky Kyiv Polytechnic Institute"
Kiev, Ukraine
liskinslava@gmail.com

Syrota Sergiy
PhD., Associated professor, Department of Applied Mathematics
National Technical University of Ukraine "Igor Sikorsky Kyiv Polytechnic Institute"
Kiev, Ukraine
sergiy.syrot@gmail.com

*Abstract* — **Based on the experience of using the "Moodle", the application of new ontology-based intelligent information technologies is proposed. In the article, proposed is a new e-learning information technology based on an ontology driven learning engine, which is matched with modern pedagogical technologies. With the help of proposed engine and developed question database we have conducted an experiment, where students were tested. The developed ontology driven system of e-learning facilitates the creation of favorable conditions for the development of personal qualities and creation of a holistic understanding of the subject area among students throughout the educational process.**

## I. INTRODUCTION

The use of new intellectual information technologies in the educational system allows to improve the learning process through introduction the new methods and approaches not only in education, but also through evaluation of acquired knowledge.

In the dawn days of WEB 2.0 development and social networks new forms of e-learning have emerged, which is owed to networked interaction between students and teachers, as well as so called collective knowledge [1].

Modern ontology-driven information systems are designed for conceptualization of ontological categories and improvement of hierarchical structures of entities on all levels. [2].

The shared use of a universal understanding of information structure by people and software agents is one of the most common goals behind the development of ontologies [3]. Lately, the OWL language is used as a standard of language for exchange between ontologies in Ontolingua, as proposed in the Semantic Web project. [4].

With the development of distance education as a form of organization of the learning process, which is characterized by provision of means for students to acquire knowledge on their own through the use of advanced informational resources that are based on modern information technologies, an issue arises with application of an individual-oriented approach in electronic education and adaptation of distance education to the individual student.

The merging of information technologies and innovative pedagogical methods is able to increase the effectiveness and quality of educational programs, increase the adaptability and individual based orientation of the educational system towards the perception and knowledge levels of those who study. At the present stage of development of education, adaptive learning systems that are based on information technologies are what is most commonly used to achieve this goal.

An important direction in the development of e-learning is the construction of cross-subject ontologies [2] based on previously developed content.

Today, testing is one of the most commonly used forms of verification of student knowledge. The use of testing in the educational process allows for short-term verification of knowledge levels for large groups of students, control of achievement of educational results and shortens data processing times. At the same time, the creation of effective and verified tests is a relatively laborious process, which involves a lot of routine work.

## II. THE CONCEPT OF THE ENGINE OF AN ONTOLOGY DRIVEN E-LEARNING SYSTEM

E-learning promotes the creation of conditions for the development of individual qualities among students during the education process. It is regarded as a means to increase the effectiveness of assimilation of learning material through audial, visual and kinesthetic types of perception. The simultaneous use of all three types of perception pushes the student towards stereoscopic perception. [5].

Therefore, the concept of an engine has to provide for the use of all three types of perception. At the same time, the problem of creation and integration of high quality content still remains on the table, and has to be resolved through the use of ontologies [6].

The concept of the engine of an ontology driven e-learning system (ODELS) is proposed in [7] and is based on a "client-server" architecture. The structure of ODELS is presented on figure 1.

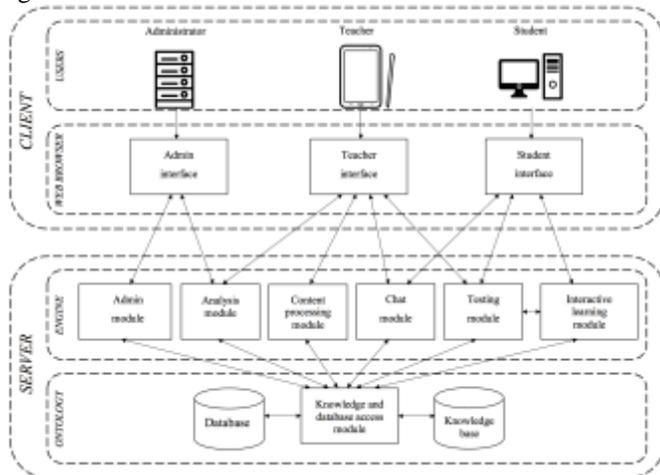

Figure 1. The structural plan of an ontology driven distance e-learning system

The proposed concept of an engine utilizes all of the advantages of e-learning systems and provides the ability to integrate developed discipline ontologies into the learning process.

### III. THE MATCHING WITH MODERN PEDAGOGICAL TECHNOLOGIES

The didactic system is a constituent of the pedagogical system and includes the following components: targeted, semantic, technological, diagnostic [8]. A compulsory condition for the implementation of the learning process is consistency on all levels.

The most well-known model that describes the learning and thinking process is Bloom's taxonomy [9].

Bloom's taxonomy is a hierarchical system of educational goals, which encompass three levels of activity: cognitive, affective, and psychomotor. According to classification by Bloom's taxonomy of the six different levels of cognitive educational goals, it encompasses the thinking process beginning with the simplest forms and ranging up to the most complicated: knowledge, comprehension, application, analysis, synthesis, evaluation. [10].

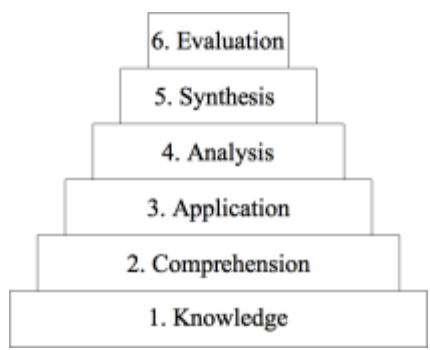

Figure 2. Levels of Bloom's taxonomy

In 2001, Lorin Anderson and his colleagues have proposed an updated version of Bloom's taxonomy, which accounts for a more wide range of factors, which influence the teaching and learning process. Just like Bloom's taxonomy, Anderson's taxonomy accounts for six levels of intellectual skills. They are arranged starting with the simplest, and ending with the most complicated: remembering, understanding, applying, analyzing, evaluating, and creating. However, the new taxonomy does not contain the synthesis level, and rather than analysis, the highest level is considered to be creating, which requires students to possess skills to generate, plan, and create something new [10, 11, 12].

The following levels have to be applied in order to facilitate the formation of higher thinking skills among students: analysis, synthesis, evaluation, and creating, which is not represented in Bloom's taxonomy, but is present in another taxonomy by L. Anderson and D. Krathwohl [10, 11].

Today, the learning system encompasses the different levels of taxonomy in the following manner:

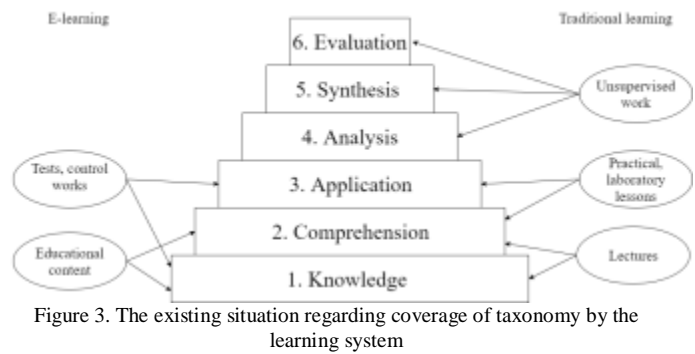

Figure 3. The existing situation regarding coverage of taxonomy by the learning system

As seen on figure 3, the last levels of the taxonomy are completely dependent on unsupervised work by the student, while the "knowledge", "comprehension", and "application" levels are shared between traditional learning and e-learning.

The following situation in coverage of taxonomy is proposed with the goal of completely covering taxonomy and building of an individual-oriented learning space:

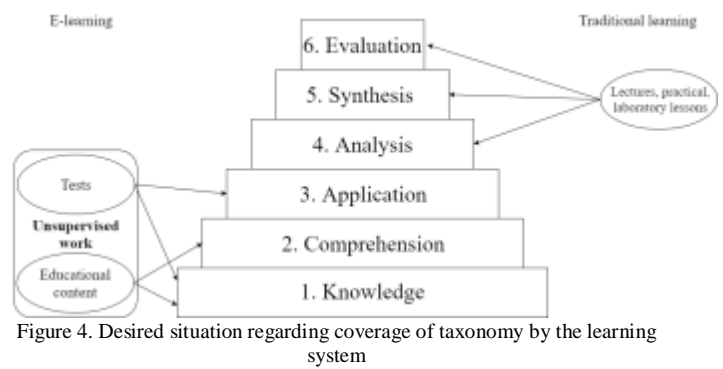

Figure 4. Desired situation regarding coverage of taxonomy by the learning system

The situation in taxonomy coverage proposed by the author of this article on figure 4, enables construction of the learning process in such a way, where the teacher would create problematic situations for comprehension and perception by students, search for resolutions of various problems through constructive interaction during lectures, practical, and laboratory lessons. Students' unsupervised work on the other hand is based on e-learning, which consists of two components: learning content and monitoring of knowledge acquisition.

The proposed redistribution of structure and educational goals for the engine of the ontology driven e-learning system will elevate the level of education quality by means of coverage of more advanced levels of cognitive thinking during lessons.

As opposed to traditional learning, e-learning has its own peculiarities: it is built on a student's independent cognitive activity and is the most individually-oriented.

TABLE I. ANALYSIS OF LEARNING PROCESS COMPONENTS FOR TRADITIONAL AND E-LEARNING

| Learning process component | Traditional learning | E-learning |
|---|---|---|
| Motivational | The motivation for learning is found in the contents of the discipline. Pursuit of good marks drives motivation. | Internal motivation to learn characteristic to each student. Improvement of motivation through interactive learning items. |
| Theoretical | Low level of thought development owed to teaching of predefined knowledge. Predominantly verbal form of teaching stimulates only auditory perception of information. | Active participation by students in the learning process. The ability to select learning topics independently accounts for individual qualities of the student. Simultaneous auditory and visual information feed, predominantly visual perception of information. |
| Practical | Template-restricted learning, practicing of skills under teacher supervision. | Independence and interactivity of the student during acquisition skills. Development of creative thinking. |
| Evaluational | Weak level of self-control. The teacher conducts the majority of evaluation and control. | High level of control. Computer software conducts the final evaluation, ruling out any bias. |

Based on table 1 it is possible to highlight the major differences between organization of traditional and e-learning:
- E-learning attracts more attention to methods of stimulation of learning motivation, than traditional learning;
- Revitalization of students' cognitive processes during e-learning is achieved through interactivity and active involvement in interaction with information technologies;
- The self-sufficiency of students while acquiring knowledge, skills and abilities is higher during e-learning as opposed to traditional learning.

This way, in order make the acquisition of learning information effective, it is necessary to account for pedagogical requirements to the introduction and use of information technologies in the modern learning process and modern educational purposes as well as psychological nuances of learning organization under the conditions of e-learning.

IV. CONSTRUCTION OF A DISCIPLINE ONTOLOGY AND GENERATION OF A QUESTION DATABASE

*A. Discipline ontology*

Based on analysis of ontological systems [4] and usage experience of the e-learning "Moodle" system [13] it is possible to resolve the highly relevant problem related to quality control of student knowledge, and in particular the development of a methodical framework for creation of ontologies of learning disciplines aimed at subsequent creation of high quality content based on the former and automation technologies for creation of test questions.

The proposed concept of meta-ontology consists two parts, content and didactic.

$$O^{Meta} = <O^{Didactic}, O^{Content}> \quad (1)$$

Based on the studied experience we can conclude that the base element (entity) of the didactic part of the meta-ontology is a set of three elements – chunk, content mapping, and relation.

$$O^{Didactic} = <Ch, L, R>, \quad (2)$$

where $Ch = \{ch_i\}$ – set of chunks which compose the didactic ontology;

$L_i = \{l_{1_i}, ..., l_{m_i}\}$ – set of content mappings;

$R = \{r_j\}$ – set of relations.

Content mappings connect content blocks within the framework of one discipline and varying types of content, whether it is a text file, presentation, video file, or test. In addition to that, content mappings serve to connect chunks with content from other disciplines.

A chunk points to a certain part of the subject area, which is provided to the student for learning. The relation between the didactic order of chunks points to chunk sequential arrangement.

In this manner, for example, a chunk from the "Algebra and Geometry" discipline called "The system of linear equations" has to follow studying of chunks with smaller significance (figure 5).

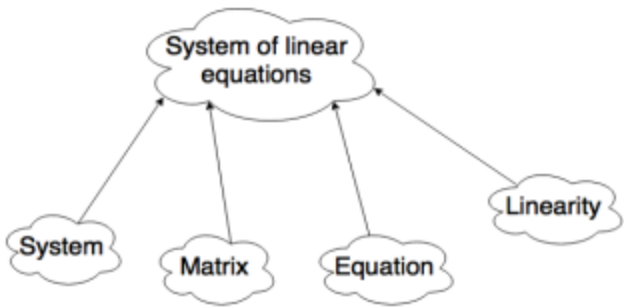

Figure 5. The "System of Linear Equations" chunk

On the other hand, the content model of meta-ontology of a learning discipline is represented with the following three items:

$$O^{Content} = <C, A, R>, \quad (3)$$

where $C = \{c_i\}$ is a set of objects which compose the ontology of the $O^{Content}$ learning discipline;

$A_i = \{a_{1_i}, \ldots, a_{n_i}\}$ is a set of properties of the $c_i$ objects, $n$ is the number of properties that describe the given object;

$R = \{r_k\}$ is the set of relations between objects and their properties.

In order to construct a content ontology it is necessary to highlight the main entities of the discipline. The given entities are matched with the following categories (figure 6). In its turn, each category has a set of certain properties [14].

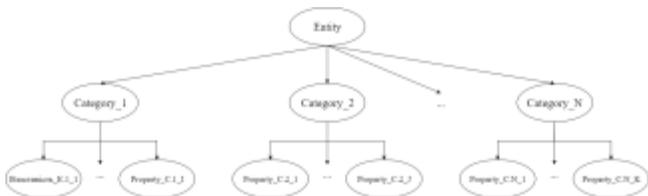

Figure 6. The principle of learning discipline content ontology construction

The set of objects, which compose the ontology of the $O^{Content}$ learning discipline coupled with the properties of these objects and the relations between them, represent the knowledge of students in the area of a certain discipline.

### B. Generation of a question database

With the help of the developed technology for generation of test questions based on ontologies and their software implementation [6,15], we have created a database of questions for disciplines such as "Data structure and Algorithms", "Computer Graphics" and "Algebra and Geometry", which are taught at the faculty of applied math of the National Technical University of Ukraine "Igor Sikorsky Kyiv Polytechnic Institute" by the authors of the given article.

Analysis of work techniques with test questions has shown that in the case of a low quantity of questions in the database it is advisable to utilize tests only for final control of acquired knowledge in the form of exams.

### C. Cross discipline referencing

The ability for cross discipline referencing of objects in electronic learning courses is one of the main features of the developed model of meta-ontology.

Due to the didactic part, the ontologies of different disciplines that are build according to a common pattern allow for automatic detection of cross discipline references, and thus creation of new cross discipline ontologies.

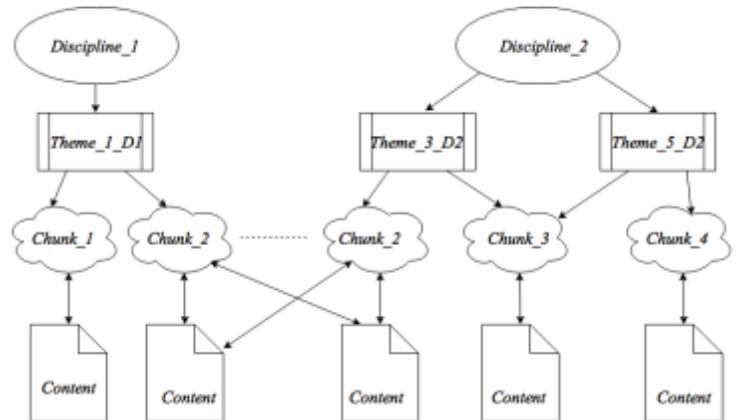

Figure 7. Cross subject referencing

For example, course tests for "Computer Graphics" include the use of mathematical chunks such as "Vector" and "Coordinate System".

If a student is unable to pass this test successfully, with the help of the didactic ontology it will be clear that the given student has not learned those chunks, and therefore has to be urged to repeat not only lectures on vector graphics, but also certain lectures on vector algebra. The said example clearly demonstrates cross subject referencing between the "Algebra and Geometry" and "Computer Graphics" courses with the help of chunks from the subject areas of "Vector" and "Coordinate System". The example that is described above is represented on figure 8.

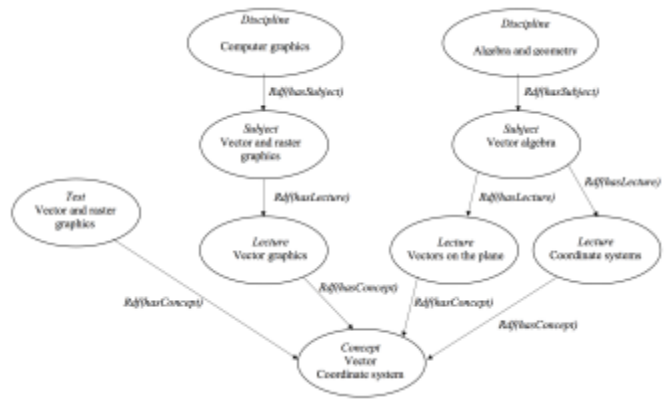

Figure 8. Cross subject connection between courses

The connection between course subjects and concepts of subject areas allows to indirectly connect lectures, tests and methodic materials between each other.

In this manner, the task for formation of a structure of the learning material in the subject area of the discipline is achieved through determination of educational content properties (theoretical and practical fragments, test question), which are based on concepts.

## V. STUDENT KNOWLEDGE TESTING AND RESULT ANALYSIS

In order for the developed automation technologies of test question and calculation task creation [6, 15] to implement the principles of individualization and differentiation of the learning process, we have matched test question types to the corresponding competences in Bloom's taxonomy (table 2).

TABLE II. COMPETENCES – QUESTION TYPE

| Competences | Test question type | Differentiation |
|---|---|---|
| Knowledge | TF - True/False questions; SA - Single answer questions | I level of difficulty |
| Comprehension | MA - Multiple Answer questions | II level of difficulty |
| Application | M, Mapping the question to its appropriate answer | III level of difficulty |

If a student submits answers to each question from level I correctly, he is eligible for a mark up to Satisfactory (E or D). This indicates that the student has acquired base skills in this discipline, and is able to resolve typical tasks that consist of one or two steps (actions).

If a student submits answers to each question from level II correctly, he is eligible for a mark up to Good (B or C). This indicates that the student is able to perform tasks correctly, which consist of two-three steps (actions), with sufficient explanation.

If a student submits answers to each question from level III correctly, he is eligible for a mark up to Excellent (A). This indicates that the student is able to perform tasks correctly, which consist of three-four steps (actions), with sufficient explanation.

With the help of the developed question database and ODELS we have conducted an experiment, where three groups of students were presented with three tests for a single topic:
– Traditional testing before ODELS use;
– Testing with ODELS in "learning" mode;
– Testing with ODELS in "exam" mode.

After the conduction of the said experiment the results of the students were analyzed as represented on figure 9.

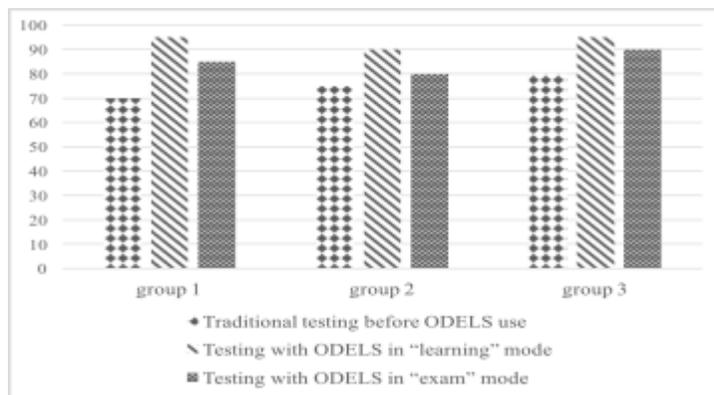

Figure 9. Results of three tests

As seen on figure 9, while testing in "learning" mode students indicate the best result, due to the system allowing reviewing information about concepts in a different window.

On the other hand, while testing in "exam" mode, students have increased their knowledge levels by at least 10% as compared to testing before ODELS implementation. This is due to higher motivation and attentiveness, as well as an individual approach to discipline learning in ODELS.

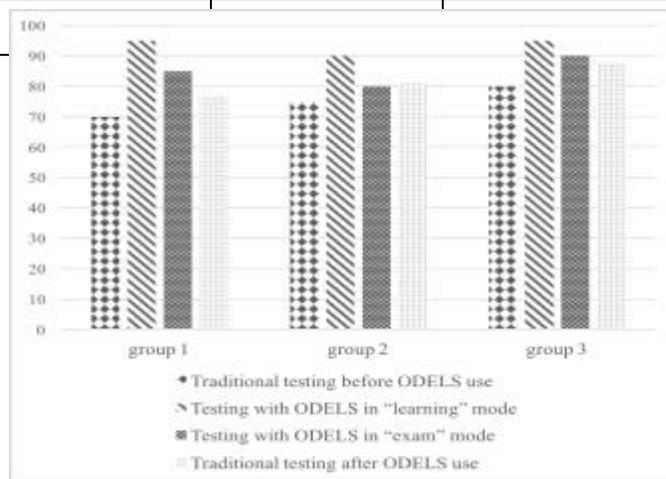

Figure 10. Additional testing results

As seen on figure 10, during repeated traditional testing after application of ODELS students have also increased their level of knowledge by at least 7% as compared to initial testing.

During analysis of the acquired testing results, we have determined that students which were proposed to undergo testing assisted by ODELS dealt with the tasks at hand with more success due to learning while working on tests. Additionally, during result analysis students that underwent level of knowledge control assisted by ODELS were easier to indicate blank spots thanks to use of the DCI (Digital concept index), which is proposed in [14].

Traditionally, while evaluating test results, any points assign for correct replies are summarized, while in the case of using DCI, points for correct replies are annulled if s student replies incorrectly to another question with the same DCI.

Such a technique allows us to rule out the possibility of a positive mark in cases of random answer guessing. During evaluation of testing results, an entry threshold is established for each DCI group, which represents course sections. For example, answers to question Q for each DCI are distributed as per table 3.

TABLE III. EXAMPLE OF EVALUATION OF A STUDENTS' WORK WITH USE OF THE DCI

|         | Q_1 | Q_2 | … | Q_m |     |
|---------|-----|-----|---|-----|-----|
| DCI_1.1 | +1  | +1  | … | +1  | 60  |
| DCI_1.2 | +1  | -1  | … | +1  | 15  |
| …       | …   | …   | … | …   |     |
| DCI_n   | +1  | -1  | … | -1  | -10 |
| Total   |     |     |   |     | 65  |

It is evident, that in the case of usage of a simple total of points, the student has achieved 65, which is a passing mark, but the entry threshold will not allow him to pass the course without additional processing by DCI_n.

Such an approach to student knowledge evaluation allows detecting of weaker points (topics) for the individual student, provide him with recommendations for repeated learning of certain sections of the discipline in the e-learning system, and stimulate him to acquiring deeper knowledge.

## VI. CONCLUSION

The developed ontology driven system of e-learning facilitates the creation of favorable conditions for the development of personal qualities and creation of a holistic understanding of the subject area among students throughout the educational process. Thanks to ODELS it is possible to increase the effectiveness of assimilation of learning material through audial, visual and kinesthetic types of perception which in turn pushes the student towards stereoscopic perception.

The results of application of the developed ontology driven system of e-learning and created database of test questions to test groups of students while teaching three disciplines, in particular "Data structure and Algorithms", "Computer Graphics" and "Algebra and Geometry", has shown that the knowledge, skills and abilities have increased by 8% on average, which brings e-learning to a new level of quality.

AUTHORS PROFILE

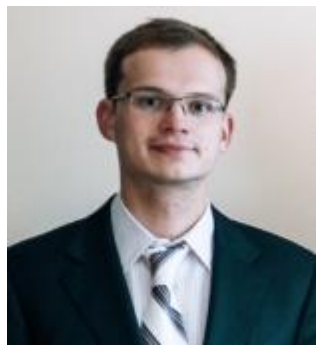

**Liskin Viacheslav,** Research student, Department of Applied Mathematics, National Technical University of Ukraine "Igor Sikorsky Kyiv Polytechnic Institute", Kiev, Ukraine. Research interests are e-learning, ontologies, machine learning, education technologies.

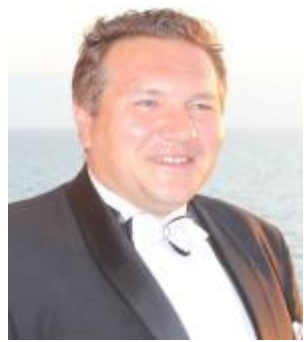

**Syrota Sergiy,** PhD., Associated professor, Department of Applied Mathematics, National Technical University of Ukraine "Igor Sikorsky Kyiv Polytechnic Institute", Kiev, Ukraine